\documentclass[12pt,a4paper]{article}
\usepackage[tbtags]{amsmath}
\usepackage{amssymb}
\usepackage{graphicx}
\usepackage[latin1]{inputenc}

\usepackage{amsbsy,amsfonts,fixmath,isomath,theorem}

\usepackage{pgf,pgfarrows,pgfnodes,tikz,color,tkz-berge}
\usetikzlibrary{arrows,shapes,snakes,automata,backgrounds,petri,topaths,calc}

\DeclareMathAlphabet{\mathonebb}{U}{bbold}{m}{n}
\newcommand{\one}{\ensuremath{\mathonebb{1}}}

\DeclareMathOperator*{\argmin}{\arg\!\min}

\newcommand{\vect}[1]{\ensuremath{  #1 } }

\newcommand{\R}{{\mathbb R}}
\newcommand{\Z}{{\mathbb Z}}

\newcommand{\D}[2]{ \ensuremath{ \frac{\mathrm{d} #1 }{\mathrm{d} #2 } }}

\newtheorem{theorem}{Theorem}[section]
\newtheorem{property}[theorem]{Property}
\begin{document}

\title{Geometric analysis of pathways dynamics: application to
versatility of TGF-$\beta$ receptors}

\author{
Satya Swarup Samal$^1$,
Aur{\'e}lien Naldi$^6$,
Dima Grigoriev$^2$, \\
Andreas Weber$^3$,
Nathalie Th{\'e}ret$^{4,5}$, and
Ovidiu Radulescu$^6$  \\
\small  $^1$ Algorithmic Bioinformatics, Bonn-Aachen International Center for IT,  Bonn, Germany, \\
\small  $^2$ CNRS, Math\'ematiques, Universit\'e de Lille,  Villeneuve d'Ascq, France, \\
\small  $^3$ Institut f{\"u}r Informatik II, University of Bonn,   Bonn, Germany, \\
\small  $^4$ Inserm UMR1085 IRSET, Universit\'e de Rennes 1, Rennes, France, \\
\small  $^5$ CNRS-Universit\'e de Rennes1-INRIA, UMR6074 IRISA, Rennes, France, \\
\small  $^6$ DIMNP UMR CNRS 5235, University of Montpellier 2, Montpellier, France.
 }

\maketitle

\centerline{\bf Abstract}

We propose a new geometric approach to describe the qualitative dynamics
of chemical reactions networks. By this method we identify
metastable regimes, defined as low dimensional regions of the phase space
close to which the dynamics is much slower
compared to the rest of the phase space. Given the network topology and
the orders of magnitude of kinetic parameters, the number of such metastable regimes is
finite. The dynamics of the network can be described as a sequence of
jumps from one metastable regime to another. We show that a geometrically computed connectivity
graph restricts the set of possible jumps. We also provide finite state machine (Markov chain) models
for such dynamic changes. Applied to signal transduction models, our approach unravels dynamical
and functional capacities of signaling pathways, as well as parameters responsible for
specificity of the  pathway response. In particular, for a model of TGF$\beta$ signalling, we find that
the ratio of TGFBR1 to TGFBR2 concentrations can be used to discriminate between metastable regimes.
Using expression data from the NCI60 panel of human tumor cell lines, we show that aggressive
and non-aggressive tumour cell lines function in different metastable regimes and
can be distinguished by measuring the relative concentrations of receptors of the two types.


\section{Introduction}
Networks of biochemical reactions are used in computational biology
as models of signaling, metabolism, and gene regulation.
For various applications it is important to understand how the dynamics of these models depend
on internal parameters, initial data and environment variables.
Traditionally, the dynamics of biochemical networks is studied in the framework of chemical kinetics
that can be either deterministic (ordinary differential equations) or stochastic (continuous time
Markov processes).
%A major inconvenience of these approaches is that parameter adjustment and validation of predictions
%require precise and complete experimental data, whereas data produced by biologists are sparse and rarely
%quantitative.
In order to cope with qualitative data, boolean or multi-valued networks are used instead of
continuous models.
Large network models have as major inconvenience the difficulty to analyse and classify their
possible dynamic behaviors. For instance, the number of states of multi-valued networks with $m$ levels
(boolean networks correspond to $m=2$) is $m^n$, which generates exponential complexity of the phase space exploration.

\begin{figure}[h!]
\begin{center}
\includegraphics[width=0.8\textwidth]{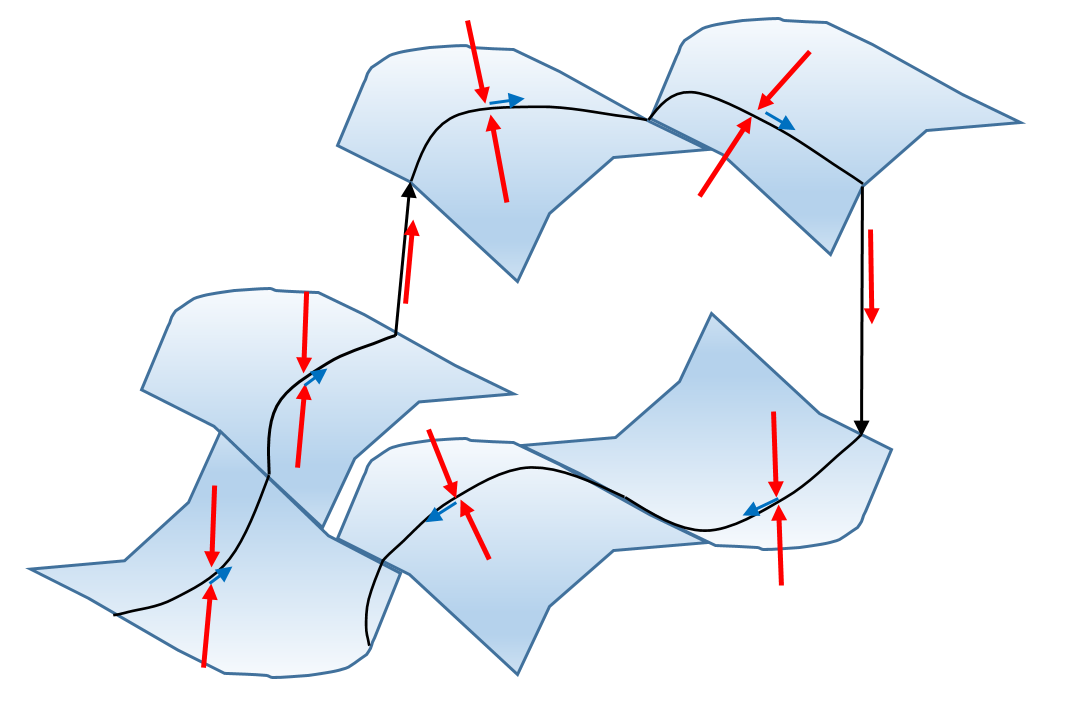}
\caption{ \label{fig:itineracy} \small
Abstract representation of metastability as itinerant trajectory in a patchy phase space landscape.
Dominant vector fields (red arrows)
confine the trajectory to low dimensional patches on which act weak uncompensated
vector fields (blue arrows). A typical trajectory contains slow segments within
patches where dominant vector fields cancel, and transitions between patches in the fast direction
of uncancelled dominant vector fields. Continuous (but non smooth) connections are also
possible, corresponding to role reversal between dominant and dominated vector fields. The term {\em crazy-quilt}
was coined to describe such a patchy landscape \cite{gorban-dynamic}.}
\end{center}
\end{figure}

In this paper we propose a new method for model analysis
that uses coarse grained descriptions of continuous dynamics as discrete
automata defined on finite states. These states will not be obtained by discretization of network variables,
but by discretization of collective  modes describing possible coordinated activity of several variables.
For large networks with
ordinary differential equations dynamics and multiple timescales
it is reasonable to consider the following property:
a typical trajectory consists in a succession of
qualitatively different slow segments separated by faster transitions.
The slow segments, generally called metastable states or regimes,
can be of several types such as attractive invariant manifolds
\cite{gorban2005invariant}, Milnor attractors \cite{rabinovich2006dynamical} or saddle connections
\cite{rabinovich2012information}.
The notion of metastability generalizes the notion of attractor.
Like in the case of attractors, distant parts of the system can have have coordinated activity for metastability.
This coordination can be abstractly represented by the proximity to a lower dimension hypersurface in the phase space.
A system remains in the proximity  of an attractor after entering its basin of attraction,
but can leave a metastable regime after a more or less long time. As illustrated in Fig.\ref{fig:itineracy}
several such transitions can happen successively.
This phenomenon, called itinerancy received particular attention
in neuroscience \cite{tsuda1991chaotic}. We believe that similar phenomena occur in molecular biology
for chemical reaction networks. A simple example sustaining this picture
is the set of bifurcations of metastable states guiding the orderly progression
of the cell cycle \cite{NGVR12sasb}. In such models, biological distinct stages
such as interphase and mitosis correspond to relatively slow segments of the dynamics and are separated
by fast transitions. During these stages, several biochemical variables
have coordinated activity.

We will use tropical geometry methods to compute metastable dynamic regimes of chemical reaction networks with polynomial
or rational kinetic laws. Tropical methods \cite{litvinov2007maslov,maclagan2009introduction},
also known as {idempotent or} max-plus algebras due
their name to the fact that one of the pioneer of the field, Imre Simon,
 was Brazilian. These methods found numerous applications to computer science \cite{simon1988recognizable},
 physics \cite{litvinov2007maslov}, railway traffic \cite{chang1998deterministic}, and  statistics \cite{pachter2004tropical}.
Recently we have applied these methods to model order reduction \cite{NGVR12sasb,Noel2013a,radulescu2015,samal2015geometric}.
In these works we have used tropical methods to rank monomial terms into rate vectors according to their
orders of magnitude and to identify lowest order, dominant terms. When there is only one dominant term
or when the dominant terms have all the same sign, the dynamics is fast and the system tends rapidly
towards a  region in phase space where at least two dominant terms of opposite signs are equilibrated.
We have called the latter situation tropical equilibration.
In this paper, we use tropical equilibrations to identify
metastable dynamic regimes of chemical reaction networks. We show that tropical equilibrations
can be grouped into branches
and describe the qualitative network dynamics as a sequence of transitions from one branch to another.
The complexity of the qualitative dynamics depends on the number of branches.
 Therefore, we would like to
know how this number depends on the number of chemical species.
Although there are theoretical results suggesting that the number of branches should be
small, these results are valid in the average in the probabilistic space of all the models.
In order to test this property numerically we will
compute the branches for a large collection of models of the Biomodels database.

The structure of the paper is as follows.
In the second section we introduce the branches of tropical equilibrations and discuss briefly
how they can be calculated.
In the third section we apply the computation of branches to models in the Biomodels database.
In the forth section we propose an algorithm to learn Markov state models defined on branches
of tropical equilibrations.
In the fifth section we apply the method to a model of TGF$\beta$ signaling
and show how the analysis can be used to interpret biological data.

\section{Tropical equilibrations of chemical reactions networks with polynomial rate functions}\label{sec:tropical}
In this section we introduce the main concepts relating geometry and dynamics.

We consider chemical reaction networks described by mass action kinetics
 \begin{equation}
 \D{x_i}{t} = \sum_j k_j S_{ij}  \vect{x}^{\vect{\alpha_{j}}}, \, 1 \leq i \leq n,
 \label{massaction}
 \end{equation}
where $k_j >0$ are kinetic constants, $x_i$ are variable concentrations, $S_{ij} \in \Z$ are the
integer entries of the stoichiometric matrix,
$\vect{\alpha}_{j} = (\alpha_1^j, \ldots, \alpha_n^j)$ are multi-indices,
  and $\vect{x^{\vect{\alpha}_{j}}}  = x_1^{\alpha_1^j} \ldots x_n^{\alpha_n^j}$, where $\alpha_i^j$ are positive integers.

For our reasonings, we can replace the exact values of parameters
by their orders of magnitude. Usually, orders of magnitude are approximations
of the parameters by integer powers of ten and serve for rough comparisons.
Our definition of orders of magnitude is based on the equation
%\begin{equation}
%\inlineequation[scaleparam]{
%k_j = \bar k_j \varepsilon^{\gamma_j}},  %\quad \gamma_j = \text{round}( \log(k_j) / \log(\varepsilon)),
%\label{scaleparam}
%\end{equation}
$k_j = \bar k_j \varepsilon^{\gamma_j}$,
where $\epsilon$ is a small positive number.
The exponents $\gamma_j$ are considered to be integer or rational. For instance, the
approximation
\begin{equation}
\gamma_j = \text{round}( \log(k_j) / \log(\varepsilon)),
\label{scaleparam}
\end{equation}
produces integer exponents,
 whereas
$\gamma_j =  \text{round}( d \log(k_j) / \log(\varepsilon)) / d$ produces rational exponents,
where round stands for the closest integer (with half-integers rounded to even numbers) and
$d$ is a strictly positive integer.
When $\epsilon = 1/10$, our definition provides the usual decimal orders.

Kinetic parameters are fixed. In contrast, species orders vary in time
and have to be computed.
To this aim, the species concentrations are first represented by orders of magnitude defined as
\begin{equation}
a_j =  \log(x_j) / \log(\varepsilon).
\label{scalespecies}
\end{equation}
Because $\log(\varepsilon) < 0$, Eq.\eqref{scalespecies} means that species orders and concentrations are
anti-correlated (large orders mean small concentrations and vice versa).

Then, network dynamics is described by a rescaled ODE system
 \begin{equation}
 \D{\bar{x}_i}{t} = \sum_j \varepsilon^{\mu_j(\vect{a}) - a_i} \bar k_j S_{ij}   {\bar{\vect{x}}}^{\vect{\alpha_{j}}},
 \label{massactionrescaled}
 \end{equation}
where
\begin{equation}
%\inlineequation[muj]{
\mu_j(\vect{a}) = \gamma_j +  \langle \vect{a},\vect{\alpha_j}\rangle ,
\label{muj}
\end{equation}
and $\langle , \rangle $ stands for the dot product. % in $\R^n$.

The r.h.s.\ of each equation in
\eqref{massactionrescaled} is a sum of multivariate monomials in the concentrations.
The orders $\mu_j$ indicate how large are these monomials, in absolute value.
A monomial of order $\mu_j$ dominates another monomial of order
$\mu_{j'}$ if
 $\mu_j < \mu_{j'}$.

To set these ideas down let us use a simple chemical network example, the Michaelis-Menten kinetics:
$$S + E \underset{k_{-1}}{ \overset{k_{1}}{\rightleftharpoons}} ES  \overset{k_2}{\rightarrow}
P + E,$$
where $S,ES,E,P$ represent the substrate, the enzyme-substrate complex, the enzyme and the product,
respectively.

After using the two conservation laws $E + ES = e_0$ and $S+ES+P= s_0$, we find
\begin{equation}\begin{split}
\label{mm}
\dot{x}_1 & = -k_1 x_1(e_0-x_2) + k_{-1} x_2 ,\\
\dot{x}_2 & = k_1 x_1 (e_0 - x_2) - (k_{-1}+k_2)x_2 .
\end{split}\end{equation}
where $x_1$, $x_2$ are the concentrations of $S$ and $ES$ respectively.

Orders of variables and parameters are as follows  $x_i= \bar x_i \epsilon^{a_i}$, $1 \leq i \leq 2$,
$k_1= \bar k_1 \epsilon^{\gamma_1}$, $k_{-1}= \bar k_{-1} \epsilon^{\gamma_{-1}}$, $e_0= \bar e_0 \epsilon^{\gamma_e}$.

{\em The tropical equilibration problem} consists in the equality of the orders of
at least two monomials one positive and another negative in the differential
equations of each species. This condition allows us to compute the concentration
orders defined by \eqref{scalespecies}.
More precisely, we want to find a vector $\vect{a}$ such that
\begin{equation}
\min_{j,S_{ij}  >0} ( \gamma_j + \langle \vect{a},\vect{\alpha_j}\rangle ) =
\min_{j,S_{ij}  <0} ( \gamma_j + \langle \vect{a},\vect{\alpha_j}\rangle )
\label{eq:minplus}
\end{equation}
The equation\eqref{eq:minplus} is related to the notion
of {\em tropical hypersurface}. A {\em tropical hypersurface} is the set of vectors
$\vect{a} \in \R^n$ such that the minimun $\min_{j,S_{ij}  \neq 0} ( \gamma_j + \langle \vect{a},\vect{\alpha_j}\rangle )$ is attained for at least two different indices $j$
(with no sign conditions). {\em Tropical prevarieties} are finite intersections of
tropical hypersurfaces. Therefore, our tropical equilibrations are subsets of
tropical prevarieties \cite{maclagan2009introduction}. The sign condition in  \eqref{eq:minplus} was imposed
because species concentrations are real positive numbers. Compensation
of a sum of positive monomials is not possible for real values of the variables.

The system \eqref{eq:minplus} can be seen as a system of equations
in min-plus algebra (also known as tropical semiring), where multiplication $\otimes$
is the real numbers addition $x \otimes y = x + y$
and the addition $\oplus$ is the minimum operation $x \oplus y = \min (x,y)$.
In order to find the solutions of such system we can explore combinatorially
trees of solutions resulting from various choices
of minimal terms and write down inequations for each situation.
Because a set of inequations define a polyhedron, the set of tropical equilibration
solutions forms  a  polyhedron in $\R^n$.
As a matter of fact, computing tropical equilibrations from the orders of magnitude of the model parameters
is a NP-complete problem \cite{theobald2006frontiers} and brute force
calculation by exploration of combinatorics has exponential complexity. However,
methods based on the Newton polytope \cite{samal2014tropical} or
constraint logic programming \cite{soliman2014constraint} exploit the
sparseness and redundancy of the system
and reduce the combinatorics and the time to compute tropical solutions.

The tropical equilibration equations for the Michaelis-Menten example are obtained by equating
minimal orders of positive monomials with minimal orders of negative monomials in \eqref{mm}:
\begin{align}
\gamma_{1}+ \gamma_{e} + a_1   &= \min (\gamma_{1} + a_1 ,\gamma_{-1} ) + a_2 ,
  \label{eq1} \\
\gamma_{1}+ \gamma_{e} + a_1  &= \min (\gamma_{1}+ a_1 , \min(\gamma_{-1},\gamma_2)  )+ a_2 .
 \label{eq2}
\end{align}

{\em Species timescales.}
The timescale of a variable $x_i$ is given by $ \frac{1}{x_i}\D{x_i}{t} = \frac{1}{\bar{x}_i}\D{\bar{x}_i}{t}$
whose order is
\begin{equation}\label{timescales}
{\nu_{i}  = \min \{ \mu_j |  S_{ij} \neq 0 \} - a_i}.
\end{equation}
The order  $\nu_{i}$ indicates how fast is the variable $x_i$ (if
$\nu_{i'} < \nu_{i}$ then $x_{i'}$ is faster than $x_{i}$) .

{\em Partial tropical equilibrations.}
It is useful to extend the tropical equilibration problem to partial equilibrations,
that means solving \eqref{eq:minplus} only for a subset of species. This is justified
by the fact that slow species do not need to be equilibrated. In order to have a
self-consistent calculation we compute the species timescales by \eqref{timescales}.
A partial equilibration is {\em consistent} if $\nu_i < \nu$ for all non-equilibrated
species $i$. $\nu > 0$ is an arbitrarily chosen threshold indicating the timescale of
interest.

{\em Tropical equilibrations, slow invariant manifolds and metastable dynamic regimes.}
In dissipative systems, fast variables relax  rapidly to some low dimensional attractive
 manifold called invariant manifold \cite{gorban2005invariant} that carries the slow mode  dynamics.
 A projection of dynamic equations
 onto this manifold provides the reduced dynamics \cite{maas1992simplifying}.
 This simple picture can be complexified to cope with hierarchies of invariant manifolds and
 with phenomena such as transverse instability, excitability and itineracy.
Firstly, the relaxation towards an attractor can have several stages, each with its own invariant manifold.
 During relaxation towards the attractor, invariant manifolds are usually embedded one into another (there is a decrease of dimensionality) \cite{chiavazzo2011adaptive}.
Secondly, invariant manifolds can lose local stability, which allow the trajectories to perform
large phase space excursions before returning in a different place on the same invariant manifold or on
a different one \cite{haller2010localized}.
We showed elsewhere that tropical equilibrations can be used to approximate
invariant manifolds for systems of polynomial differential
equations \cite{NGVR12sasb,Noel2013a,radulescu2015}.
Indeed,
tropical equilibration are defined by the cancelling out of dominant forces acting on the system. The remaining
weak non-compensated forces ensure the slow dynamics on the invariant manifold.
Tropical equilibrations are thus different from steady states, in that there is a slow dynamics.
In this paper we will use them as proxies for metastable dynamic regimes.

 More precisely, let us assume that species timescales (defined by Eq.\eqref{timescales}) satisfy the relation
$\nu_1 \leq \nu_2 \leq \ldots \leq \nu_n$ and that not all the timescales are the same,  i.e.
there is  $m < n$ such that $\nu_{m+1} - \nu_{m}  > 0$. Then,
two groups of variables have separated timescales.
The variables $\vect{X}_r = (x_1,x_2,\ldots,x_m)$ are fast (change significantly on
timescales of order of magnitude $\varepsilon^{-\nu_{m}}$ or shorter.
The remaining variables $\vect{X}_s = (x_{m+1},x_{m+2},\ldots,x_n)$ are
much slower (have little variation on timescales of order of magnitude $\varepsilon^{-\nu_{m}}$).
The metastable regime means that fast variables have reached quasi-steady state values
on a low dimensional hypersurface of the phase space.

{\em Branches of tropical equilibrations and connectivity graph.}
For each equation $i$, let us define
\begin{equation}
M_i(\vect{a}) = \underset{j}{\argmin}
 (\mu_j(\vect{a}), S_{ij}  >0) = \underset{j}{\argmin} (\mu_j(\vect{a}),S_{ij}  <0),
 \label{Mi}
\end{equation}
in other words
$M_i$ denotes the set of monomials having the same minimal order $\mu_i$.
We call {\em tropically truncated system} the system obtained by pruning the system
\eqref{massactionrescaled}, i.e. by keeping only the dominating monomials.
 \begin{equation}
 \D{\bar{x}_i}{t} = \varepsilon^{\mu_i - a_i} (\sum_{j\in M_i (\vect{a})} \bar k_j  \nu_{ji}  {\bar{\vect{x}}}^{\vect{\alpha_{j}}}),
 \label{massactionrescaledtruncated}
 \end{equation}
The tropical truncated system is uniquely determined by the index sets
$M_i(\vect{a})$, therefore by the
tropical equilibration
$\vect{a}$. Reciprocally, two tropical equilibrations can have the same index sets
$M_i(\vect{a})$ and truncated systems. We say that two tropical equilibrations $\vect{a}_1$, $\vect{a}_2$
are  equivalent iff $M_i(\vect{a}_1) = M_i(\vect{a}_2), \text{for all } i$. Equivalence classes
of tropical equilibrations
are called
{\em branches}. A branch $B$ with an index set $M_i$  is {\em minimal} if
$M'_i \subset M_i$ for all $i$ where $M'_i$ is the index set $B'$ implies
$B'=B$ or $B'=\emptyset$.
%Because for each index $i$, the relation
% \eqref{lines} defines a hyperplane, the tropical equilibration branches are on intersections of $n$
% such hyperplanes.
Closures of equilibration branches are defined by a finite set of
linear inequalities, which means that they are polyhedral complexes.
Minimal branches correspond to maximal
dimension faces of the polyhedral complex.
The incidence relations between the maximal dimension faces ($n-1$ dimensional faces,
where $n$ is the number of variables) of the polyhedral complex define
the {\em connectivity graph}. More precisely, minimal branches are the vertices of this graph.
Two minimal branches are connected if the corresponding faces
of the polyhedral complex share a $n-2$ dimensional face. In terms of index sets,
two minimal branches with index sets $M$  and $M'$ are connected if there is
an index set $M''$ such that $M'_i \subset M''_i$ and $M_i \subset M''_i$ for all $i$.

Returning to the Michaelis-Menten example let us analyse the  quasi-equilibrium situation
\cite{meiske1978approximate,segel1988validity,segel1989quasi,GRZ10ces,gorban2011michaelis}
when
the reaction constant $k_{-1}$ is much faster than the reaction constant $k_2$. In
terms of orders, this condition reads $\gamma_{-1} < \gamma_2$.
In this case, the two tropical equilibration equations \eqref{eq1}, \eqref{eq2} are identical,
because $\min(\gamma_{-1},\gamma_2) = \gamma_{-1}$.
Let $\gamma_m = \gamma_{-1} - \gamma_1$ denote the order of the parameter $K_m = k_{-1}/k_1$.
There are two branches of solutions of \eqref{eq1}, namely
$a_2 = \gamma_{e}, a_1 \leq \gamma_m $ and
$a_2 = a_1 + \gamma_{e} - \gamma_{m}, a_1 \geq \gamma_m$
corresponding
to
 $\min (\gamma_{1} + a_1 ,\gamma_{-1} ) = \gamma_{1} + a_1 $ and to
 $\min (\gamma_{1} + a_1 ,\gamma_{-1} ) = \gamma_{-1}$, respectively.
Using the relation between orders and concentrations we identify the first branch of solutions with
the saturation regime $x_2 \approx e_0$ (the free enzyme is negligible) and
$x_1 >> K_m$ (the substrate has large concentration)
and the second branch with the linear regime $x_2 << e_0$ (the
concentration of the
attached enzyme is negligible) and $x_1 << K_m$ (the substrate has low concentration).

The fast truncated system (obtained after removing all dominated monomials from \eqref{mm}) reads
\begin{equation}\begin{split}
\label{mmtrunc}
\dot{x}_1 & = -k_1 x_1 e_0  + k_{-1} x_2 , \\
\dot{x}_2 & =  k_1 x_1 e_0  - k_{-1}x_2 ,
\end{split}\end{equation}
for the linear regime branch and
\begin{equation}\begin{split}
\label{mmtruncs}
\dot{x}_1 & = -k_1 x_1(e_0-x_2), \\
\dot{x}_2 & =  k_1 x_1(e_0-x_2),
\end{split}\end{equation}
for the saturated regime branch.

\section{Benchmarking on Biomodels database}
\subsection{Data source}
For benchmarking, we selected $34$ models  from the r25 version of Biomodels database \cite{Novere2006} with polynomial vector field.

\subsection{Computation of minimal branches}

The model files are parsed and the polynomial vector fields are extracted. Thereafter, the conservation laws (that are the sum of the variables whose total concentration is invariant) are computed. The vector field along with the conservation laws are the input to the tropical geometry based algorithm in \cite{SamalGrigorievRadulescu2015a} to compute the minimal branches. It should be noted here that due to the conservation laws the number of equations may exceed the number of chemical species.

According to the Eq.\eqref{eq:minplus} and to the geometric interpretation of
tropical equilibrations from Sect.\ref{sec:tropical} the tropical solutions are either isolated
points or bounded or unbounded polyhedra. Changing the parameter $\varepsilon$
is just a way to approximate the position of these points and polyhedra by
lattices or in other words by integer coefficients vectors.
Finding the value of $\varepsilon$ that provides
the best approximation is a complicated problem in Diophantine approximation. For that
reason, we preferred an experimental approach consisting in
choosing several values of $\varepsilon$ and checking the robustness of the results.

\begin{figure}[h!]
\begin{center}
\includegraphics[width=0.8\textwidth]{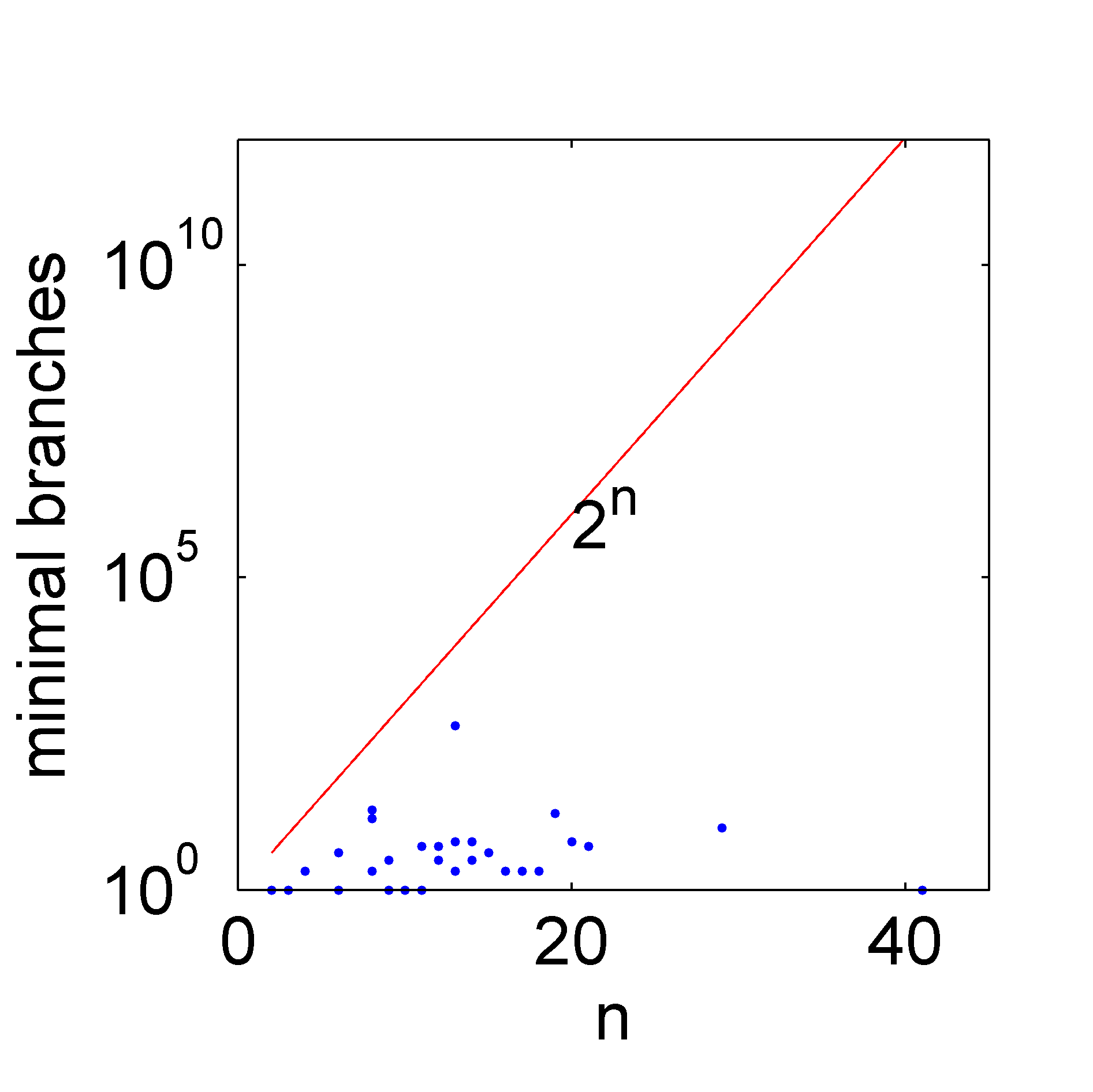}
\caption{ \label{fig:running-time} \small
Semi-log plot of the minimal branches versus the number of equations in the models from Biomodels for $\varepsilon=1/5$.
Comparison with a binary network number of states $2^n$ suggests sub-exponential scaling. }
\end{center}
\end{figure}

A summary of the analysis is presented in Table \ref{tab:Summary-of-analysis} with four different choices for $\varepsilon$ values (we wanted to have orders of magnitude close to decimal ones and to avoid commensurability between different values of  $\varepsilon$; the choice $1/5$, $1/7$, $1/9$, $1/23$ seemed good enough for this purpose).
In addition, in Fig. \ref{fig:running-time}  we plot the
 number of minimal branches versus the number of equations in the model. As can be noticed, this number is much lower than the number of states of a boolean network with the same number of variables,
which illustrates the advantage of our coarse graining with respect
to boolean or multi-value networks.

\begin{table}[h!]
\caption{\small Summary of analysis on Biomodels database\label{tab:Summary-of-analysis}.
The benchmarked models have a number of variables from 3 to 41.
Model BIOMD0000000289 has tropical branches at $\varepsilon$ values $1/5$, $1/7$, $1/9$ but none at $1/23$. Model BIOMD0000000108 has no branches at all values of $\varepsilon$ considered for benchmarking. }

\begin{small}
\begin{center}
\begin{tabular}{|c|r|r|r|r|r|r|}
\hline
\multicolumn{1}{|p{0.7cm}|}{$\varepsilon$ value} &
\multicolumn{1}{|p{1.2cm}|}{Total models con\-sidered }&
%\multicolumn{1}{|p{1.2cm}|}{{Timed-out models}}&
\multicolumn{1}{|p{1.5cm}|}{{Models without tropical branches}} &
\multicolumn{1}{|p{1.5cm}|}{{Models with tropical branches}} &
\multicolumn{1}{|p{1.5cm}|}{ Average running time (in secs)}&
\multicolumn{1}{|p{1.5cm}|}{ Average number of minimal branches}
%\multicolumn{1}{|p{1.5cm}|}{ Models with Unit-definition}

\tabularnewline
\hline
\hline
1/5 & 34 & 1 & 33 & 299.24 & 3.24\tabularnewline
\hline
1/7 & 34  & 1 & 33 & 243.69 & 3\tabularnewline
\hline
1/9 & 34  & 1 & 33 & 309.39 & 3.75\tabularnewline
\hline
1/23 & 34  & 2 & 32 & 3178.21 & 3.84\tabularnewline
\hline
\end{tabular}
\end{center}
\end{small}
\end{table}

\section{Learning a finite state machine from a nonlinear biochemical network}
In order to coarse grain the continuous model onto a finite-state automaton, we
first need a way to map the phase space of the continuous model to a finite
set of branches. First, we compute the branches of tropical solutions as subsets
of the euclidian space $\R^n$ where $n$ is the number of variables.
We are using the algorithm based on constraint solving
introduced in \cite{soliman2014constraint} to
obtain all rational tropical equilibration solutions $\vect{a} = (a_1,a_2,\ldots,a_n)$
within a box $|a_i| < b$, $b >0$ and with denominators smaller than a fixed value
$d$, $a_i = p_i/q$, $p_i,q$ are positive integers, $q < d$.
The output of the algorithm is a matrix containing all the tropical equilibrations within
the defined bounds. A post-processing treatment is applied to this output
consisting in computing truncated systems, index sets, and minimal branches.
Tropical equilibrations minimal branches are stored as matrices $A_1,A_2,\ldots,A_b$, whose lines are
tropical solutions within the same branch. Here $b$ is the number of minimal branches.

Our method computes numerical approximations of the tropical prevariety. Given a value of
$\epsilon$, this approximation is better when the denominator bound $d$ is high. At fixed
$d$, the dependence of the precision on $\epsilon$ follows more intricate rules dictated by
Diophantine approximations. For this reason, we systematically test
that the number $b$ and the truncated systems corresponding
to minimal branches are robust when changing the value of $\epsilon$.

Trajectories $\vect{x}(t) = (x_1(t),\ldots,x_n(t))$ of the smooth dynamical system are
generated with different initial conditions, chosen uniformly and satisfying the conservation laws, if any.
For each time $t$, we compute the Euclidian distance
$d_i(t) = \min_{\vect{y} \in A_i}  \left \lVert \vect{y} - log_{\varepsilon}(\vect{x}(t)) \right \rVert,$
where
$\left \lVert * \right \rVert$ denotes the Euclidean norm and
$\log_\varepsilon(\vect{x}) = (\log x_1 / \log(\varepsilon) ,\ldots, \log x_n / \log(\varepsilon))$.
This distance classifies all points of the
trajectory as belonging to a tropical minimal branch. The result is a symbolic trajectory
$s_1,s_2,\ldots$ where the symbols $s_i$ belong to the set of minimal branches.
In order to include the possibility of transition regions we include an unique symbol $t$
to represent the situations when the minimal distance is larger than a fixed threshold.
We also store the residence times $\tau_1,\tau_2,\ldots$ that represent the
time spent in each of the state.

The stochastic automaton is learned as a homogenous, finite states, continuous time Markov process,
defined by the lifetime (mean sojourn time) of each state $T_i$, $1 \leq i\leq b$ and
by the transition probabilities $p_{i,j}$ from a state $i$ to another state $j$.
We use the following estimators for the lifetimes and for the transition probabilities:
\begin{eqnarray}
T_i  &=& (\sum_{n} \tau_n  \one_{s_n = i}) / (\sum_{n} \one_{s_n = i}) \\
p_{i,j} &=& (\sum_{n} \one_{ s_n = i, s_{n+1} = j} ) / (\sum_{n}  \one_{s_n = i}), \, i \neq j
\end{eqnarray}

\section{Application to TGF-$\beta$ signalling}
\begin{figure}[h!]
\begin{center}
\scalebox{0.6}{
\vspace*{-5mm}
\begin{tikzpicture}
 \SetUpEdge[lw         = 0.5pt,
            color      = black,
            labelstyle = {sloped,scale=2}]
  \tikzset{node distance = 1.5cm}
  \GraphInit[vstyle=Normal]
  \SetVertexMath
       \tikzset{VertexStyle/.style={scale=1.0,
       draw,
            shape = circle,
            line width = 1pt,
            color = black,
            outer sep=1pt}}
  \Vertex[x=0,y=0]{B1}
  \Vertex[x=3,y=0]{B2}
  \Vertex[x=6,y=0]{B3}
  \Vertex[x=0,y=3]{B4}
  \Vertex[x=3,y=3]{B5}
  \Vertex[x=6,y=3]{B6}
    \Vertex[x=0,y=6]{B7}
  \Vertex[x=3,y=6]{B8}
  \Vertex[x=6,y=6]{B9}

  \Edge(B1)(B2)
  \Edge(B2)(B3)
    \Edge(B4)(B5)
  \Edge(B5)(B6)
    \Edge(B7)(B8)
  \Edge(B8)(B9)
    \Edge(B1)(B4)
  \Edge(B2)(B5)
    \Edge(B3)(B6)
  \Edge(B4)(B7)
    \Edge(B5)(B8)
  \Edge(B6)(B9)
    \Edge(B1)(B5)
  \Edge(B2)(B4)
  \Edge(B3)(B5)
  \Edge(B2)(B6)
    \Edge(B4)(B8)
  \Edge(B5)(B7)
    \Edge(B6)(B8)
  \Edge(B5)(B9)
\draw(3,-1.5)node[above,right, scale =2] {a)};

\begin{scope}[xshift=10cm]
\SetUpEdge[lw         = 0.5pt,
            color      = black,
            labelstyle = {sloped,scale=2}]
  \tikzset{node distance = 1.5cm}
  \GraphInit[vstyle=Normal]
  \SetVertexMath
       \tikzset{VertexStyle/.style={scale=1.0,
       draw,
            shape = circle,
            line width = 1pt,
            color = black,
            outer sep=1pt}}
  \Vertex[x=0,y=0]{B1}
  \Vertex[x=3,y=0]{B2}
  \Vertex[x=6,y=0]{B3}
    \Vertex[x=0,y=3]{B4}
  \Vertex[x=3,y=3]{B5}
  \Vertex[x=6,y=3]{B6}
    \Vertex[x=0,y=6]{B7}
  \Vertex[x=3,y=6]{B8}
  \Vertex[x=6,y=6]{B9}
\tikzset{EdgeStyle/.style={post,line width = 0.5, font=\tiny}}
\Edge[label=$0.96$](B2)(B1)
\Edge[label=$0.03$](B5)(B1)
\Edge[label=$0.42$](B5)(B4)
\Edge[label=$0.01$](B5)(B3)
\Edge[label=$0.29$](B5)(B6)
\Edge[label=$0.16$](B8)(B9)
\Edge[label=$0.11$](B8)(B4)
\Edge[label=$0.1$](B8)(B6)
\Edge[label=$0.53$](B8)(B5)
\Edge[label=$0.1$](B8)(B7)
\Edge[label=$1.0$](B7)(B4)
\Edge[label=$1.0$](B9)(B6)
\path (B1) edge [loop below] node {0.999} (B1);

\tikzset{EdgeStyle/.style={post,bend right,line width = 0.5, font=\tiny}}
\Edge[label=$0.0004$](B1)(B4)
\Edge[label=$1.0$](B4)(B1)
\Edge[label=$0.005$](B2)(B5)
\Edge[label=$0.24$](B5)(B2)
\Edge[label=$0.0004$](B3)(B6)
\Edge[label=$1.0$](B6)(B3)
\Edge[label=$0.04$](B2)(B3)
\Edge[label=$0.999$](B3)(B2)

\draw(3,-1.5)node[above,right, scale =2] {b)};
\end{scope}
  \end{tikzpicture}
}
\vspace*{-5mm}
\begin{tabular}[b]{cc}
 \scalebox{0.33}{\includegraphics[width=19cm]{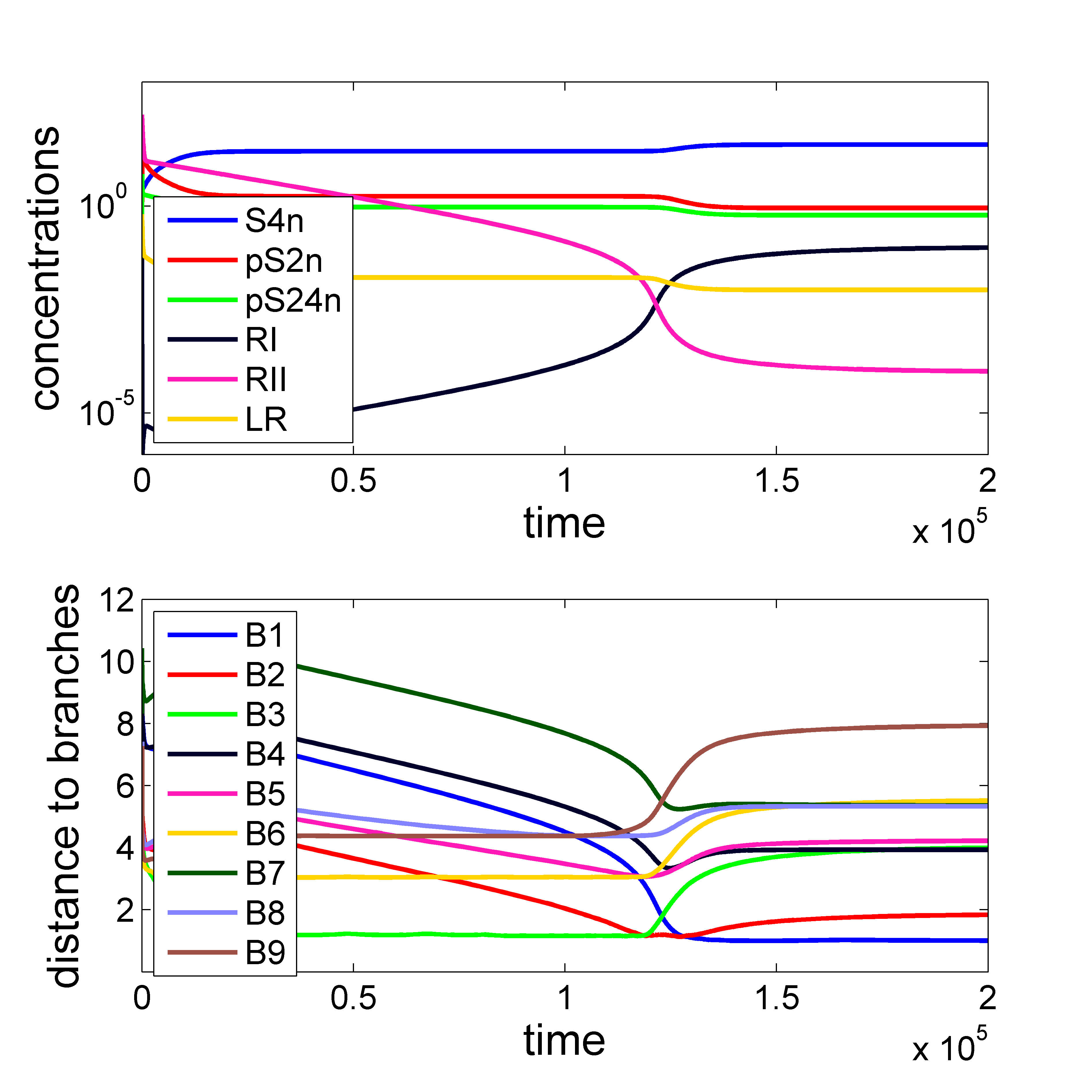}} &  \scalebox{0.33}{\includegraphics[width=19cm]{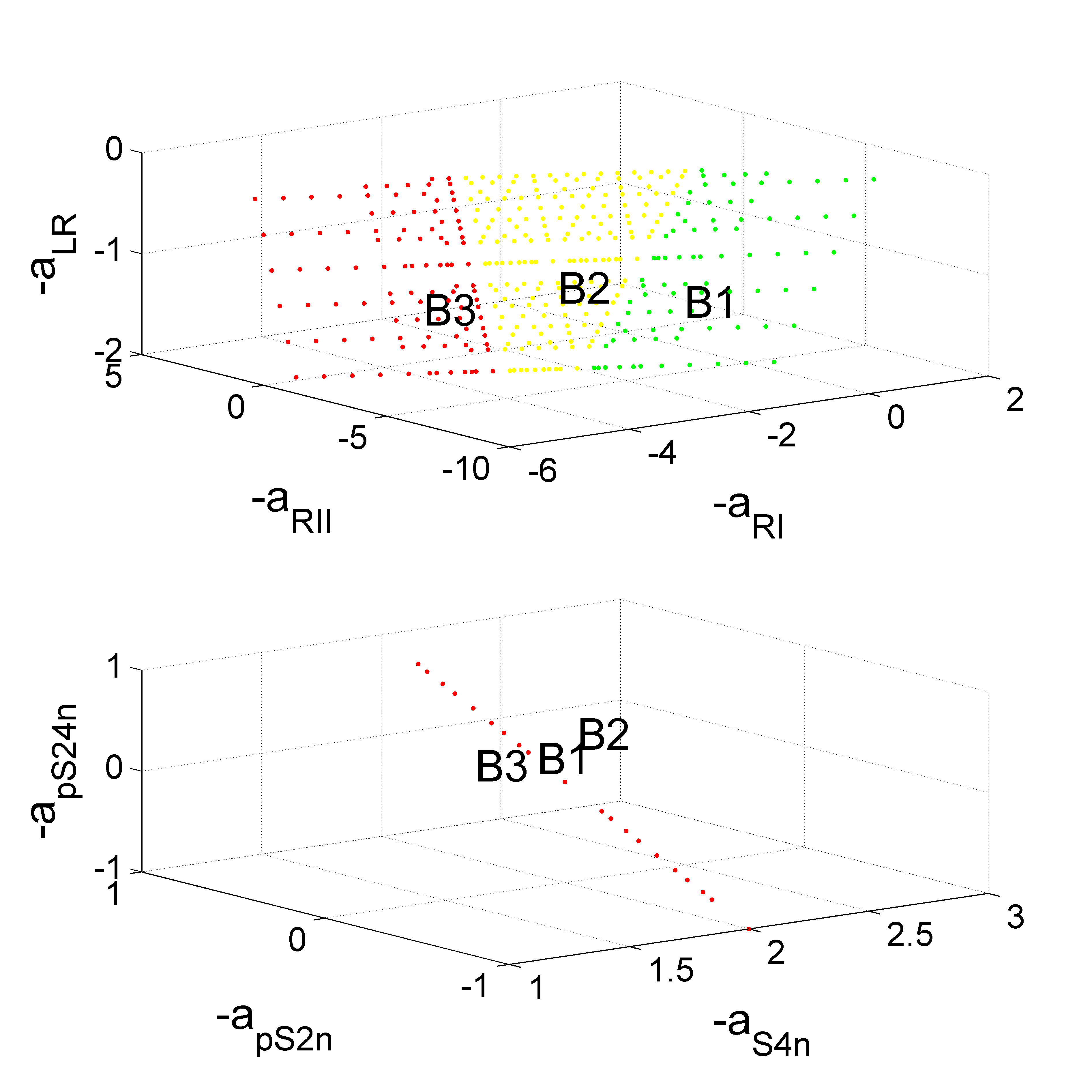}} \\
 c) & d)
\end{tabular}
\end{center}
\caption{\label{example2}
TGF$\beta$ model. a) Connectivity graph of tropical minimal branches; b) finite
state automaton; c) a single trajectory of the system (starting from initial
data chosen randomly close to the branch $B_3$) is represented by plotting the
concentration of different species vs. time (upper sub-figure);
the distances to different branches of solutions vs. time (lower sub-figure)
shows that the sequence of branches for this trajectory
is $B_3$, $B_2$, $B_1$ (all points of the trajectory are close
to one of these three branches and significantly more distant
to the other branches).
d) The different branches of solutions are defined by allowed
concentrations of different variables, represented here by orders of magnitudes $a_i$;
the opposite concentration orders $-a_i$ are
proportional to  the logarithms of concentrations $-a_i \sim \log(x_i)$.
The most used branches are $B_1$, $B_2$, $B_3$ are shown in projection onto sets of
three variables.
The variables RI, RII, LR are plasma membrane receptors
and ligand-receptor complex
(signaling input layers), whereas pS2n, S4n, pS24n are nuclear transcription factors
and complexes (effectors). The structure of tropical branches shows that composition of
input layers is more flexible (varies on planar domains that are disjoint for different branches)
than the concentrations of effectors (vary on linear intervals that overlap for different branches). }
\end{figure}

As a case study we consider a nonlinear model of dynamic regulation of
Transforming Growth Factor beta TGF-$\beta$ signalling pathway that we have recently described
in \cite{andrieux2012dynamic}. TGF-$\beta$ signalling occurs through association with a
heteromeric complex of two types of transmembrane serine/threonine kinases, the type I (TGFBR1)
and type II (TGFBR2) receptors.  TGF-$\beta$  binding to TGFBR2 induces recruitment and phosphorylation
of TGFBR1, which in turn transmits the signal through phosphorylation of
Smad2 transcription factor. Once phosphorylated, the Smad2 hetero-dimerizes
with Smad4 and the complexes then migrate to the nucleus, where they regulate
the transcription of TGF-$\beta$-target genes. In that context, the
Transcriptional Intermediary Factor 1, TIF1-$\gamma$ have been shown
to function either as a transcriptional repressor or as an alternative
transcription factor that promote TGF-$\beta$ signaling.
The apparent controversial effect of TIF1-$\gamma$ on regulation of the Smad-dependent
TGF-$\beta$ signalling was solved by a model integrating a ternary complex associating
TIF1-$\gamma$ with Smad2 and Smad4 complexes.
This model has a dynamics defined
by $n=18$ polynomial differential equations and  $25$ biochemical reactions.
The computation of the tropical equilibrations for this model shows that
there are 9 minimal branches of full equilibrations (in these
tropical solutions all variables are equilibrated).
The connectivity graph of these branches  and the learned
finite-state automaton are shown in Figure~\ref{example2}.

The transition probabilities of the automaton are coarse grained properties of the statistical ensemble of trajectories for different initial conditions. Given a state and a minimal branch close to it, it will depend on the actual trajectory to which other branch the system will be close to next. However, when initial data and the full trajectory are not known, the automaton will provide estimates of where we go next and with which probability.
For the example studied and for nominal parameter values, the branch B1 is a globally attractive sink: starting from anywhere, the automaton will reach B1 with probability one. This branch contains the unique stable steady state of the initial model. This calculation illustrates the basic properties of minimal branches of equilibrations.
Trajectories of the dynamical system can be decomposed into segments that remain close
to minimal branches. Furthermore, all the observed transitions between branches are contained in the connectivity graph resulting from the polyhedral complex of the tropical equilibration branches.
The connectivity graph can be thus used to constrain the possible transitions.
A change of parameter values can have several consequences: change the connectivity graph, change of
the probabilities of transitions and change of the attractor position.

\begin{figure}[h!]
\begin{center}
\scalebox{1}{
\begin{tikzpicture}
 \SetUpEdge[lw         = 0.5pt,
            color      = black,
            labelstyle = {fill=white, sloped}]
  \tikzset{node distance = 6cm,main node/.style={circle,fill=blue!12,draw,font=\sffamily\large\bfseries}
   }
 \GraphInit[vstyle= normal]
  \SetGraphUnit{2}
\node[main node] (x12){x12};
\node[main node] (x13)[right of=x12]{x13};
\node [main node](x15)[below of=x12,node distance = 4cm] {x15};
\node [main node](x16)[below of=x13,node distance = 4cm] {x16};
\path[] (x12) to node[midway, main node](x14){x14} (x16) ;
\node [main node](x11)[below of=x14,node distance = 2cm] {x11};
\node [](x12l)[left of=x12,node distance = 3cm] {};
\node [](x13r)[right of=x13,node distance = 3cm] {};
\node [](x12bl)[below left of=x12,node distance = 3cm] {};
\node [](x13br)[below right of=x13,node distance = 3cm] {};
\path[] (x14) to node[midway](x14br){} (x16) ;
\path[] (x14) to node[midway](x14bl){} (x15) ;
\tikzstyle{EdgeStyle}=[post,line width = 1]
\Edge[label=$k_{20}$](x12)(x12bl)
\Edge[label=$k_{18}$](x12l)(x12)
\Edge[label=$k_{19}$](x13r)(x13)
\Edge[label=$k_{21}$](x13)(x13br)
\Edge[label=$k_{23}$](x14)(x11)
\tikzstyle{EdgeStyle}=[post,bend left,line width = 1]
\Edge[label=$k_{24}$](x14)(x14br)
\Edge[label=$k_{26}$](x12)(x15)
\Edge[label=$k_{27}$](x15)(x12)
\Edge[label=$k_{28}$](x13)(x16)
\Edge[label=$k_{29}$](x16)(x13)

\coordinate (O1) at ($(x12)!0.5!(x13)$);
\coordinate[label=$k_{22}k_{35}$] (O2) at ($(O1)!0.25!(x14)$);
\tikzstyle{EdgeStyle}=[bend right,line width = 1]
\Edge[](x13)(O2)
\tikzstyle{EdgeStyle}=[bend left,line width = 1]
\Edge[](x12)(O2)
\tikzstyle{EdgeStyle}=[post,line width = 1]
\Edge[](O2)(x14)

\coordinate (O3) at ($(x12)!0.4!(x14)$);
\tikzstyle{EdgeStyle}=[bend left,line width = 1]
\Edge[label=$k_{30}$](x11)(O3)
\tikzstyle{EdgeStyle}=[post, bend right,line width = 1]
\Edge[](O3)(x12)
\tikzstyle{EdgeStyle}=[post, bend left,line width = 1]
\Edge[](O3)(x13)

\tikzstyle{EdgeStyle}=[post,bend right,line width = 1]
\Edge[label=$k_{25}$](x14)(x14br)

\end{tikzpicture}
}
\centerline{\bf Full model}
\scalebox{1}{
\begin{tikzpicture}
 \SetUpEdge[lw         = 0.5pt,
            color      = black,
            labelstyle = {fill=white, sloped}]
  \tikzset{node distance = 6cm,main node/.style={circle,fill=blue!12,draw,font=\sffamily\large\bfseries}
   }
 \GraphInit[vstyle= normal]
  \SetGraphUnit{2}
\node[main node] (x14){x14};
\node[main node] (x13)[right of=x14, node distance = 3cm]{x13};
\node[main node] (x12)[left of=x14, node distance = 3cm]{x12};

\node [](O2)[above of=x14,node distance = 2cm] {};
\coordinate [label=$k_{22}k_{35}$] (OO2) at (O2);
\tikzstyle{EdgeStyle}=[bend right,line width = 1]
\Edge[](x13)(OO2)
\tikzstyle{EdgeStyle}=[bend left,line width = 1]
\Edge[](x12)(OO2)
\tikzstyle{EdgeStyle}=[post,line width = 1]
\Edge[](OO2)(x14)

\node [](O3)[below of=x14,node distance = 2cm] {};
\coordinate (OO3) at (O3);
\tikzstyle{EdgeStyle}=[line width = 1]
\Edge[label=$k_{23}$](x14)(OO3)
\tikzstyle{EdgeStyle}=[post,bend left,line width = 1]
\Edge[](OO3)(x12)
\tikzstyle{EdgeStyle}=[post,bend right, line width = 1]
\Edge[](OO3)(x13)

\node [](x12l)[left of=x12,node distance = 2cm] {};
\node [](x13r)[right of=x13,node distance = 2cm] {};
\tikzstyle{EdgeStyle}=[post,line width = 1]
\Edge[label=$k_{18}$](x12l)(x12)
\Edge[label=$k_{19}$](x13r)(x13)

\path (x12) edge [loop below, line width = 1] node {$k_{26}$} (x12);
\path (x13) edge [loop below, line width = 1] node {$k_{28}$} (x13);

\coordinate (O1) at ($(x14)!0.5!(x13)$);
\tikzstyle{EdgeStyle}=[post,bend left,line width = 1]
\Edge[label=$k_{24}$](x14)(O1)
\tikzstyle{EdgeStyle}=[post,bend right,line width = 1]
\Edge[label=$k_{25}$](x14)(O1)

\end{tikzpicture}
}
\centerline{\bf Reduced model}
\end{center}
\caption{Graphic representation of the ligand-receptor module of the TGF-$\beta$
full model.
A reduced model
%(valid if $\gamma_{26} < \gamma_{20}$, $\gamma_{28} < \gamma_{21}$
%meaning that internalization is more rapid than degradation for both
%receptors 1 and 2)
was found (see Appendix 2) that has the same tropical solutions as the full model
and is used to simplify the calculation of the branches of
tropical equilibrations.
The different variables mean:
$x_{12}:\, \text{RI}$ (TGBR1),
$x_{13}:\, \text{RII}$ (TGFBR2),
$x_{14}:\, \text{LR}$ (ligand-receptor complex),
$x_{15}:\, \text{RIe}$ (TGFBR1 in endosome),
$x_{16}:\, \text{RIIe}$ (TGFBR2 in endosome),
$x_{11}:\, \text{LRe}$ (LR in endosome).
}
\label{fig:receptors}
\end{figure}
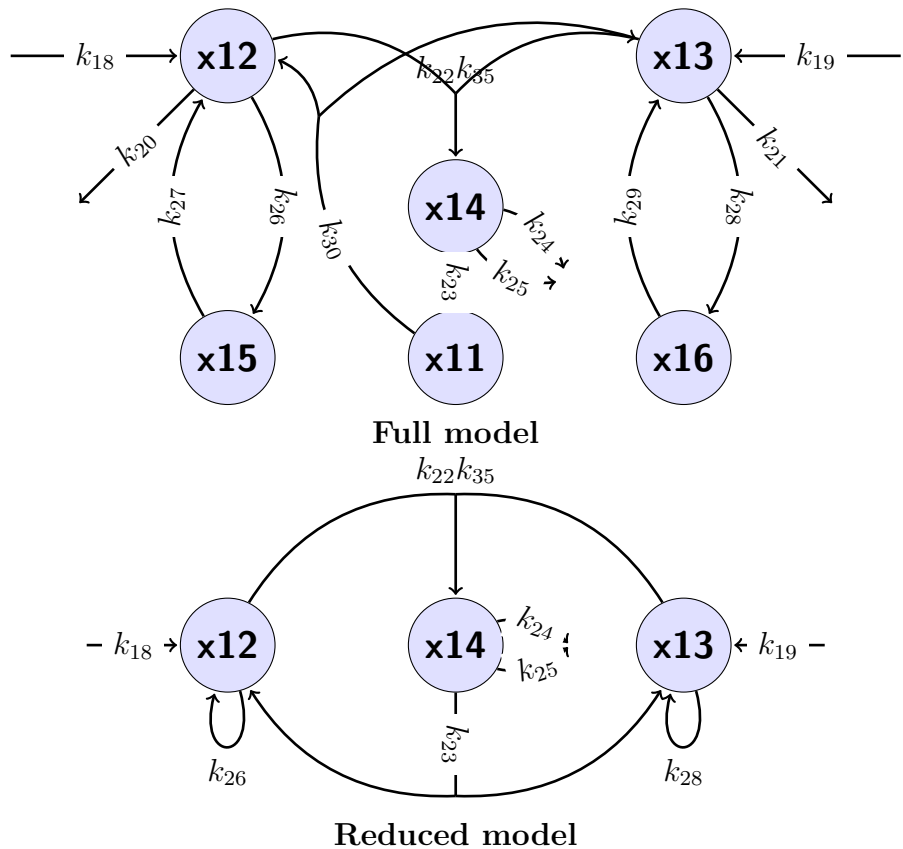

In order to understand the significance of the minimal branches
and their relation with dynamic and
physiologic properties of the network we have performed an analytic study of the tropical
equilibration solutions. We show in the following that the most important cause of the multiplicity of
branches is the dynamics of the TGFBR1 and TGFBR2 receptors whose internalization
and trafficking regulates TGF-$\beta$ signalling \cite{le2005clathrin}. These two receptors belong to a ligand-receptor module of 6 variables and 12 reactions that is decoupled from the rest of the
network. More precisely, the ligand-receptor module activates the SMAD transcription factors
but receives no feed-back (see Figure~\ref{fig:receptors}) and can be studied independently from the rest of the
variables. This module has been used with little variation in many models of
TGF-$\beta$ signalling \cite{vilar2006signal,zi2007constraint,chung2009quantitative}.

We show in the appendix 2 that the tropical equilibration of the ligand-receptor module
form a two dimensional polyhedron conveniently parametrized by the concentration orders $a_{12}$
and $a_{13}$ of the receptors TGFBR1 and TGFBR2 respectively. The branches can be calculated
analytically (Eqs.\eqref{pol},\eqref{eq:branches}). For the nominal values of the model parameters one of these branches is empty and the three remaining branches correspond to $B_1$, $B_2$ and $B_3$.
The two other triplets of
branches $(B_4, B_5, B_6)$ and  $(B_7, B_8, B_9)$ correspond to the same mutual relations of
variables in the ligand-receptor module. They are distinguished by the values of the remaining
variables (the transcription factors module). Our computation of the automaton showed that
the branches $B_i,\, i \in [4,9]$ are practically inaccessible from states in
branches $B_i,\, \, i \in  [1,3]$, therefore we will not discuss them here.
%If the condition
%$\min (\gamma_{24},\gamma_{25}, \gamma_{23})  = \gamma_{23}$ (meaning that complex
%decomposition  plasma membrane ligand-complex loss, being at least as
%fast as complex degradation)
%is fulfilled, then the polyhedron of solution is
%\begin{equation}
%\begin{split}
%{\mathcal P} = & (\{ a_{12}  + a_{13} + \gamma_{22} + \gamma_{35} \leq \gamma_{18} \}
%\cup
%\{ \gamma_{26}  + a_{12} \leq \gamma_{18} \}) \cap \\
%& (\{ a_{12} + a_{13} + \gamma_{22} + \gamma_{35} \leq \gamma_{19} \}
%\cup  \{ \gamma_{28} + a_{13} \leq \gamma_{19} \}).
%\end{split}
%\end{equation}
%The polyhedron of solutions
%can be decomposed into at most 4 minimal branches defined
%by one of the conditions
%\begin{eqnarray}
%\{ a_{12} + \gamma_{26} < a_{12} + a_{13} + \gamma_{22} + \gamma_{35} \} \cap
%\{ a_{13} + \gamma_{28} < a_{12} + a_{13} + \gamma_{22} + \gamma_{35} \}  \\
%\{ a_{12} + \gamma_{26} < a_{12} + a_{13} + \gamma_{22} + \gamma_{35} \} \cap
%\{ a_{13} + \gamma_{28} > a_{12} + a_{13} + \gamma_{22} + \gamma_{35} \}  \\
%\{ a_{12} + \gamma_{26} > a_{12} + a_{13} + \gamma_{22} + \gamma_{35} \} \cap
%\{ a_{13} + \gamma_{28} < a_{12} + a_{13} + \gamma_{22} + \gamma_{35} \} \\
%\{ a_{12} + \gamma_{26} > a_{12} + a_{13} + \gamma_{22} + \gamma_{35} \}
% \cap  \{ a_{13} + \gamma_{28} > a_{12} + a_{13} + \gamma_{22} + \gamma_{35} \}
% \end{eqnarray}

\begin{figure}[h!]
\begin{center}
\begin{tabular}[b]{cc}
 \scalebox{0.33}{\includegraphics[width=18cm]{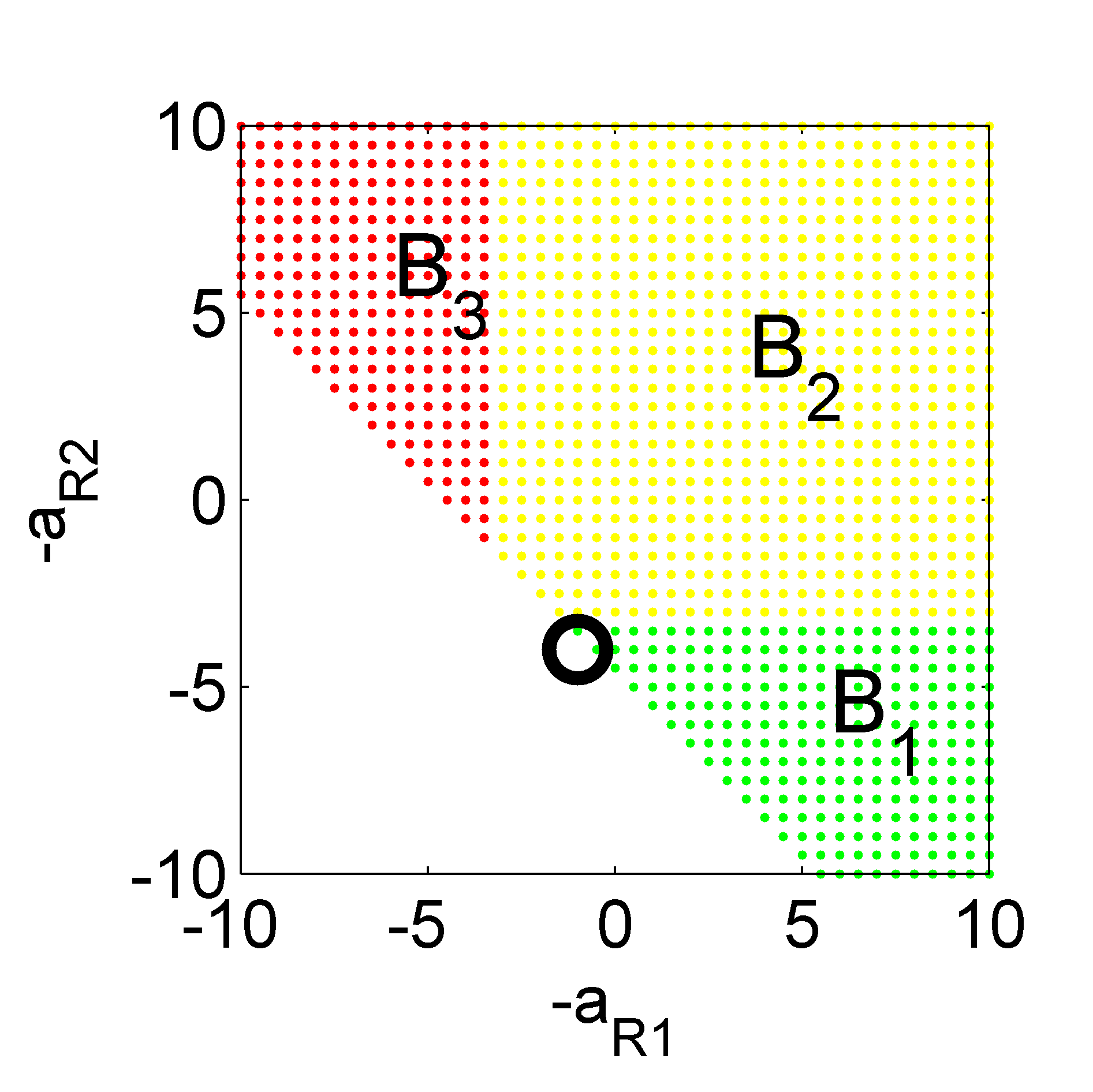}} &  
 \scalebox{0.33}{\includegraphics[width=18cm]{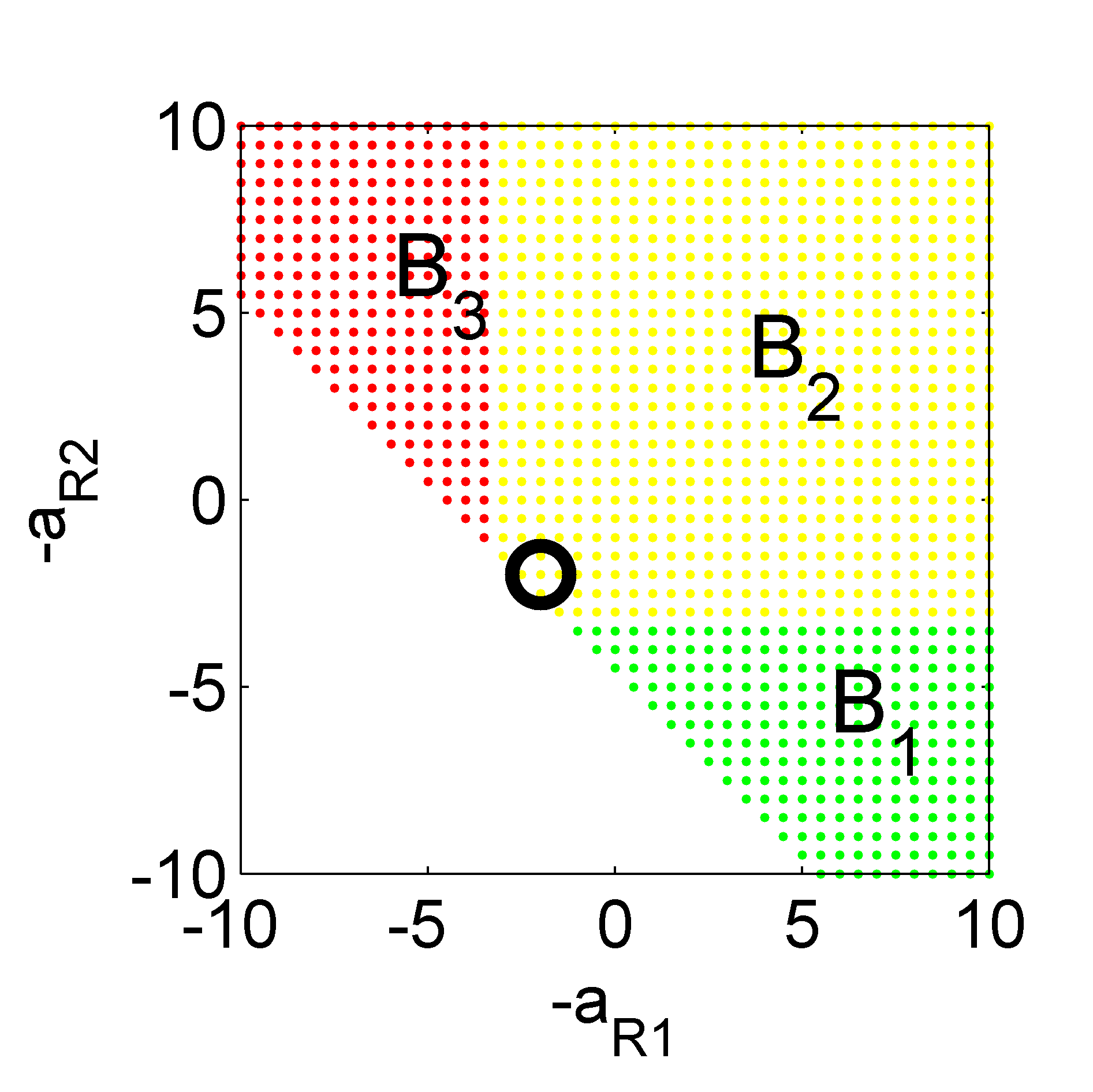}} \\
 a) & b) \\
 \scalebox{0.33}{\includegraphics[width=18cm]{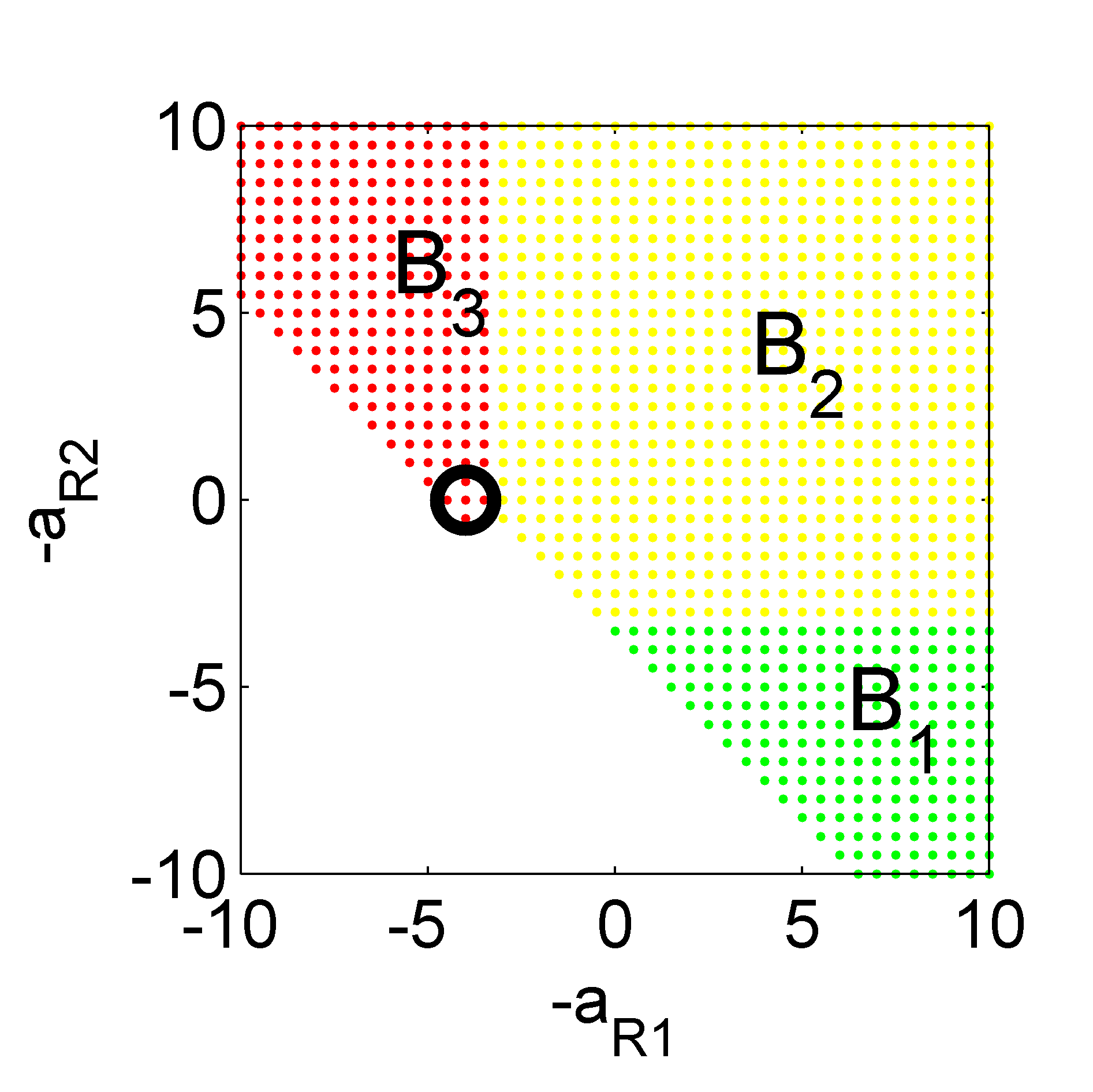}} &
 \scalebox{0.33}{\includegraphics[width=20cm]{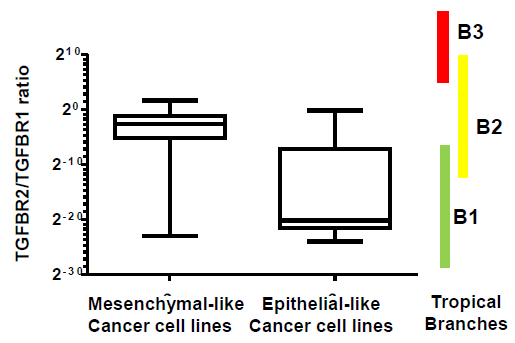}} \\
  c) & d)
\end{tabular}
\end{center}
\caption{\label{branches}
Tropical equilibrations of the ligand-receptor modules for various values of TGFBR2 (R2) gene expression
are represented in projection in the plane $(a_{R1},a_{R2})$ (a point in this plane provides the
orders of concentrations of the protein receptors)
(a,b, and c) and comparison with proteomics data from \cite{gholami2013global} (d).
Branches of tropical equilibrations are calculated
for a) nominal value $k_{19}$ (TGFBR2 expression) (this is the same as
Fig.\ref{fig:receptors}d in projection onto the plane $(a_{R1},a_{R2})$),
b) $\times 2$  TGFBR2  overexpression , and c) $\times 10$ TGFBR2 overexpression. The circle represents
the position of the stable steady state and the branch containing is an attractor of the finite-state
automaton.
d) Proteomic data from NCI-60 cancer cell lines. Aggressive lines cover
concentration domains corresponding to branches 3 and 2, whereas non-aggressive lines
correspond to low expression of TGFBR2 in branch 1.
}
\end{figure}

We have used symbolic computation to determine the steady states of the ligand-receptor module.
This module has an unique steady state corresponding to concentrations orders that can be placed inside the
polyhedron of tropical solutions using the Eq.\eqref{scalespecies}.
The minimal branch containing the steady state is a sink of the coarse grained dynamics.
The polyhedron of tropical solutions, its decomposition into minimal branches,
and the position of the
steady states inside it, depend on model parameters.
Among model parameters two are important: $k_{18}$ and $k_{19}$ representing
the production rate of the protein receptors TGFBR1 and TGFBR2, respectively.
Consequently, these two parameters are correlated to gene expression and account for
possible variability in mRNA levels of the two types of receptors.
Figure~\ref{branches} shows the tropical equilibration branches of the ligand-receptor modules
for various parameters $k_{19}$ corresponding to various TGFBR2 expression levels.
For the nominal parameters used in the model, the branch $B1$ is an attractor (the coarse-grained dynamics shows that the probability to leave this state is negligible), and the branches $B2$ and $B3$ are only metastable. This means that starting in the branch $B2$ or $B3$
the receptor module will reach the branch $B1$ after a certain time and will remain there.
However, over-expression of TGFBR2 modelled by changing the parameter $k_{19}$ and illustrated
in the Figure~\ref{branches}a-c can tilt the balance in favor of large concentrations
of receptor of type 2 corresponding to branches $B2$, $B3$ of tropical solutions in the model.
% and to
%aggressive tumors in the data.
Interestingly, this change occurs by a displacement of the position of the steady state
from $B1$ to $B2$ and $B3$ and not by a change of the concentration values allowed for
these branches.

While Vilar et al \cite{vilar2006signal} have speculated that the ligand-receptor module
is responsible for the versatility of the response of the TGF-$\beta$ pathway, no experimental evidence support this hypothesis.
Here, we now demonstrate that there are correlations
between dynamical specificity characterized by membership to a particular branch of equilibration
and cell phenotype.
We illustrate such a comparison for the NCI-60 panel of cancer cell lines,
a well established tool for tumour comparison and drug screening provided by the National
Cancer Institute. Based on
microarray analysis,
these cell lines were found to cluster into two classes: epithelium-like (non-aggressive) and mesenchymal-like (aggressive) cell lines  \cite{ross2000systematic}.

Using the global proteome analysis of this NCI-60 panel \cite{gholami2013global}
we extracted the protein expression levels of TGFBR1 and TGFBR2 and showed that
mesenchymal-like (aggressive) cell lines can be distinguished from epithelial-like (non-aggressive)
cell lines by the increased level of TGFBR2 (Fig~\ref{branches}d.

The proteome data was compared to membership to tropical branches.
According to Eq.\eqref{scalespecies} there is a linear relation between opposite
concentration orders $-a_i$ and logarithms of
concentrations, $-a_i \approx  b \log (x_i),\, i =1,\ldots,n$ ($b = - 1/\log(\varepsilon) > 0$).
We used opposite concentration orders $-a_i$ instead of $a_i$ because they change in the same direction
as the concentrations (small opposite orders mean small concentrations and large opposite
orders mean large concentrations). Therefore, in Fig~\ref{branches}a-c the relation
$TGFBR1=TGFBR2$ is verified on the bissector of the first quadrant, whereas
$TGFBR2>TGFBR1$ and $TGFBR2<TGFBR1$ are valid above and below the bissector, respectively.
%a non-zero
%intercept, the same for all variables $x_i,\, i =1,\ldots,n$, should be introduced in this linear relation to cope
%with concentration units; however, the non-zero intercept would shift the entire graph along the
%bisector $TGFBR1=TGFBR2$ with no effect on the fold ratios of receptor concentrations), and consequently can be qualitatively compared to proteome data.
When compared the proteome results with the membership to a particular branch of equilibration, we found that the distribution of concentration orders in branches  place non-aggressive cancer cell lines  in a range
covered by branch 1, whereas the aggressive cancer cell lines are placed in a range covered by
branches 2 and  3 (Fig~\ref{branches}d. Indeed, the ratio $TGFBR2/TGFBR1$ is small for the branch
$B_1$ and in non-aggressive cancer cell lines, and is much larger for $B_2, B_3$ and in
aggressive cell lines.

Furthermore, we validated the association of up-regulation of TGFBR2 with mesenchymal-like appearance in an independent dataset of 51 breast cancer cell lines \cite{neve2006collection}. As we have recently described \cite{ruff2015disintegrin}, comparative analyses between Basal B cell lines with mesenchymal-like phenotype and  Basal A and Luminal cell lines with epithelial morphology permitted to identify more than 600 differentially expressed genes that include TGFBR2. Gene expression data were now extracted for TGFBR1 and TGFBR2 and  we showed that TGFBR2 gene expression is significantly induced in mesenchymal-like cell lines while TGFBR1 did not vary (Supplementary Figure 1).
In accordance with our observation, Parker et al. \cite{parker2007lower} have previously reported the association of low TGFBR2 expression with a lower aggressive tumour phenotype.

 In the present work, our tropical approach has identified TGFBR2 concentration as a discriminant parameter for defining metastable regimes. The importance of such up-regulation of TGFBR2 in aggressive cancer cell  lines might be related to its implication in Smad-independent signaling that includes PI3K-Akt, JNK, p38MAPK and Rho-like GTPases  and which highly contribute to epithelial-mesenchymal transition
\cite{zhang2009non,moustakas2012induction}.

Together these observations suggest that metastable regimes defined by branches of minimal tropical equilibrations are associated with cell phenotypes. The idea of associating tropical minimal branches with clinical phenotype is similar to the idea of cancer attractors
\cite{Huang2009869} where the idea is that cancer cells are trapped
in some abnormal attractors.

%Figure~\ref{example2} bottom right shows the structure of most probable branches, the ones in which the systems spends most of his time. The branches B1, B3 and B2 correspond to different compositions of the membrane and of the endosome, rich in the receptor RI, rich in the receptor RII and rich in both types of receptors, respectively. Even if this
%compositon is changed on wide domains of orders (planes in the space of orders), the concentrations
%of effectors are robust (are more constained than the concentrations of receptors).

\section{Conclusion}
We have presented a method to coarse grain the dynamics of a smooth biochemical reaction network
to a discrete symbolic dynamics of a finite state automaton. The coarse graining was obtained
using a tropical geometry approach to compute the states. These states correspond to metastable
dynamic regimes and to relatively slow segments of the system trajectories.
The coarse grained model can be used for studying statistical properties of biochemical
networks such as occurrence and stability of temporal patterns, recurrence, periodicity and
attainability  problems.

Further improvement and evolution is possible for this approach. First, the
coarse graining can be performed in a hierarchical way. For the nonlinear
example studied in the paper we computed only the full tropical equilibrations that
stand for the lowest order in the hierarchy (coarsest model). As discussed in Section~3
we can also consider partial equilibrations when slow variables are not equilibrated and thus refine
the automaton. Generally, there are more partial equilibrations than total equilibrations and
learning an automaton on the augmented state set will produce refinements. Second, and most
importantly, the dynamics within a branch could be also described. As shown elsewhere,
reductions of the systems of ordinary equations are valid locally close to
tropical equilibrations \cite{NGVR12sasb,Noel2013a,radulescu2015,samal2015geometric}. Furthermore, the same reduction is valid for all the equilibrations
in a branch. This suggests that a hybrid approach, combining reduced ODE dynamics within
branch with discrete transitions between branches is feasible. The transitions can be
autonomously and deterministically commanded by crossing the boundaries between branches
that are perfectly determined by our approach.

The most important result of this paper is the extension of the notion of attractor to
metastable regimes of chemical reaction networks and the proposition of a practical
recipe to compute metastability.
Metastable regimes correspond to low-dimensional hypersurfaces of the phase space, along
which the dynamics is relatively slower.
Most likely, metastable regimes have biological
importance because the network spends most of its time in these states. The itinerancy of the network,
described as the possibility of transitions from one metastable regime to another is paramount
to the way neural networks compute, retrieve and use information \cite{tsuda1991chaotic}
and can have similar role in biochemical networks.
Our approach based on tropical geometry provides an algorithmic method
 to test these ideas further. The extension of this approach i.e. making use of statistical methods to compute the association of the tropical minimal branches with clinical phenotypes based on ``-omics'' data remains a topic of future research.

%The idea of associating tropical minimal branches with clinical phenotype is similar to the idea of cancer attractors \cite{Huang2009869} where the idea is that cancer cells are trapped in some abnormal attractors. Our approach based on tropical geometry provides an algorithmic method to test it further. We demonstrated that the non-aggressive tumour is associated with branch 1, whereas the aggressive tumour is associated with branches 2 and 3. The extension of this approach i.e. making use of statistical methods to compute the association of the tropical minimal branches with clinical phenotypes based on {\em omics} data remains a topic of future research.

\paragraph{Acknowledgements}
O.R and A.N are supported by INCa/Plan Cancer grant N$^\circ$ASC14021FSA.

%\bibliographystyle{plain}
%\bibliography{references_trop_unified}

%\newpage
%\setcounter{page}{1}

\paragraph{Appendix 1: Description of the TGFb model used in this paper.}

The model is described by the following system of differential equations
\begin{eqnarray}
\D{x_1}{t}&=& 	 k_{2} x_{2} - k_{1} x_{1} - k_{16} x_{1} x_{11}\notag \\
\D{x_2}{t}&= &	 	 k_{1} x_{1} - k_{2} x_{2} + k_{17} k_{34} x_{6}\notag \\
\D{x_3}{t}&= &	 	 k_{3} x_{4} - k_{3} x_{3} + k_{7} x_{7} + k_{33} k_{37} x_{18} -k_{6} x_{3} x_{5}\notag \\
\D{x_4}{t}&= &	 	 k_{3} x_{3} - k_{3} x_{4} + k_{9} x_{8} - k_{8} x_{4} x_{6}\notag \\
\D{x_5}{t}&= &	 	 k_{5} x_{6} - k_{4} x_{5} + k_{7} x_{7} + 2 k_{11} x_{9} - 2 k_{10} x_{5}^2 - k_{6} x_{3} x_{5} + k_{16} x_{1} x_{11}\notag \\
\D{x_6}{t}&= &	 	 k_{4} x_{5} - k_{5} x_{6} + k_{9} x_{8} + 2 k_{13} x_{10} - 2 k_{12} x_{6}^2 - k_{17} k_{34} x_{6} + k_{31} k_{36} x_{8} - k_{8} x_{4} x_{6}\notag \\
\D{x_7}{t}&= &	 	 k_{6} x_{3} x_{5} - x_{7} (k_{7} + k_{14})\notag \\
\D{x_8}{t}&= &	 	 k_{14} x_{7} - k_{9} x_{8} - k_{31} k_{36} x_{8} + k_{8} x_{4} x_{6}\notag \\
\D{x_9}{t}&= &	 	 k_{10} x_{5}^2 - x_{9} (k_{11} + k_{15})\notag \\
\D{x_{10}}{t}&=& 	 	 k_{15} x_{9} - k_{13} x_{10} + k_{12} x_{6}^2\notag \\
\D{x_{11}}{t}&= &	 	 k_{23} x_{14} - k_{30} x_{11}\notag \\
\D{x_{12}}{t}&= &	 	 k_{18} - x_{12} (k_{20} + k_{26}) + k_{30} x_{11} + k_{27} x_{15} - k_{22} k_{35} x_{12} x_{13}\notag \\
\D{x_{13}}{t}&= &	 	 k_{19} - x_{13} (k_{21} + k_{28}) + k_{30} x_{11} + k_{29} x_{16} - k_{22} k_{35} x_{12} x_{13}\notag \\
\D{x_{14}}{t}&= &	 	 k_{22} k_{35} x_{12} x_{13} - x_{14} (k_{23} + k_{24} + k_{25}) \notag \\
\D{x_{15}}{t}&= &	 	 k_{26} x_{12} - k_{27} x_{15}\notag \\
\D{x_{16}}{t}&= &	 	 k_{28} x_{13} - k_{29} x_{16}\notag \\
\D{x_{17}}{t}&= &	 	 k_{31} k_{36} x_{8} - k_{32} x_{17}\notag \\
\D{x_{18}}{t}&= &	 	 k_{32} x_{17} - k_{33} k_{37} x_{18}\label{system}
\end{eqnarray}

These variables are as follows:
\begin{itemize}
\item
Receptors on plasma membrane: $x_{12}=$ RI (receptor 1), $x_{13}=$ RII (receptor 2), $x_{14}=$ LR (ligand-receptor complex).
\item
Receptors in the endosome: $x_{11}=$ LRe, $x_{15}=$ RIe, $x_{16}=$ RIIe.
\item
Transcription factors and complexes in cytosol: $x_1=$ S2c,  $x_3=$ S4c, $x_5=$ pS2c, $x_7=$ pS24c, $x_9=$ pS22c, $x_{18}=$ S4ubc.
\item
Transcription factors and complexes in the nucleus: $x_2=$ S2n, $x_4=$ S4n, $x_6=$ pS2n, $x_8=$ pS24n, $x_{10}=$ pS22n, $x_{17}=$ S4ubn.
\end{itemize}

\paragraph{Appendix 2: Calculation of tropical equilibration branches for the TGFb model used in this paper.}

Tropical equilibration solutions for the variables $x_{11}$,$x_{12}$, $x_{13}$, $x_{14}$, $x_{15}$, $x_{16}$ can be computed
independently from the rest of the variables. The ordinary differential equations for these variables
form a subsystem that is decoupled (receives no feed-back) from the rest of the equations.

We can reduce the system of 6 tropical equations to a simplified system of 3 tropical equations using the following
two general properties.

\begin{property}[binomial species] \label{prop1}
$Y$ is a binomial species if the ordinary differential equation defining its rate of variation contains
only one positive monomial term and only one negative monomial term
$$\D{Y}{t} = M_1(\vect{X}) Y^{n_1} - M_2(\vect{X}) Y^{n_2},$$
where $\vect{X}$ denotes the other variables. We further assume that $n_1 < n_2$.
Then, the species $Y$ can be eliminated and the resulting simplified tropical system has the
same tropical equilibration solutions as the full system. The simplification is performed by eliminating the equation for $Y$
and replacing everywhere $Y$ by $(M_1/M_2)^{1/(n_2-n_1)}$.
\end{property}

%\begin{pop1}
The proof follows from the fact that the tropical equation for the order $a$ of $Y$ is linear in $a$ (there is
no min operation) and therefore it has the unique solution $a = \frac{1}{(n_2-n_1)}(\mu_1 - \mu_2)$.
%\end{pop1}

\begin{property}[dominated first order reactions] \label{prop2}
If a species $Y$ is consumed by several first order reactions of kinetic constants $k_1, k_2, \ldots, k_r$
and if  $\gamma_1 \leq \gamma_2 \leq \ldots \leq \gamma_p < \gamma_{p+1} \leq \gamma_{p+2} \leq \ldots \leq \gamma_r$, then
the reactions $k_{p+1},\ldots, k_r$ can be eliminated and the resulting simplified tropical system has the
same tropical equilibration solutions as the full system.
\end{property}

%\begin{pop2}
The proof follows from the following obvious property of the min operation $\min ( \gamma_1, \ldots, \gamma_p, \ldots, \gamma_r ) =
\min ( \gamma_1, \ldots, \gamma_p)$.
%\end{pop2}

Using $\gamma_{26} < \gamma_{20}$, $\gamma_{28} < \gamma_{21}$
(a condition satisfied by the nominal model parameters and meaning that internalization is more rapid than degradation for both
receptors 1 and 2) and the Properties \ref{prop1},\ref{prop2}
we can justify the reduction illustrated in Figure \ref{fig:receptors}.
Because the reduced model has the same tropical solutions as the full, larger model,
it is enough to solve the tropical equilibration problem for the reduced model. This reads
\begin{eqnarray}
\min(\gamma_{18},a_{14}+\gamma_{23}, a_{12} + \gamma_{26}) & = &
\min (a_{12} + a_{13} + \gamma_{22} + \gamma_{35}, a_{12} + \gamma_{26} ) \label{teq1} \\
\min (\gamma_{19}, a_{14} + \gamma_{23}, a_{13} + \gamma_{28}) &=&
\min (a_{12} + a_{13} + \gamma_{22} + \gamma_{35}, \gamma_{28} + a_{13})
\label{teq2} \\
\min (\gamma_{24},\gamma_{25}, \gamma_{23}) + a_{14} &=&
a_{12} + a_{13} + \gamma_{22} + \gamma_{35} \label{teq3}
\end{eqnarray}
Suppose now that the following condition is true
\begin{equation}
\min (\gamma_{24},\gamma_{25}, \gamma_{23})  = \gamma_{23}.
\end{equation}
This condition is satisfied by the nominal parameters and, like the previous condition, means that receptors
have relatively large life-times.
Then from \eqref{teq3} we got
$a_{14} = a_{12} + a_{13} + \gamma_{22} + \gamma_{35} - \gamma_{23}$
and
the equations \eqref{teq1},\eqref{teq2} become
\begin{eqnarray}
\min(\gamma_{18},a_{14}+\gamma_{23}, a_{12} + \gamma_{26}) & = &
\min (a_{14} + \gamma_{23}, a_{12} + \gamma_{26} ) \label{teq1bis} \\
\min (\gamma_{19}, a_{14} + \gamma_{23}, a_{13} + \gamma_{28}) &=&
\min (a_{14} + \gamma_{23},  a_{13} + \gamma_{28})
\label{teq2bis}
\end{eqnarray}
The solutions of \eqref{teq1bis}, \eqref{teq2bis} can be easily found
and form the following polyhedron
\begin{eqnarray}
(\{ a_{12}  + a_{13} + \gamma_{22} + \gamma_{35} \leq \gamma_{18} \}
&\cup& \{ \gamma_{26} + a_{12} \leq \gamma_{18} \})  \cap \notag \\
(\{ a_{12} + a_{13} + \gamma_{22} + \gamma_{35} \leq \gamma_{19} \}
&\cup&  \{ \gamma_{28} + a_{13} \leq \gamma_{19} \}), \notag \\
a_{14} &= &a_{12} + a_{13} + \gamma_{22} + \gamma_{35} - \gamma_{23}. \label{pol}
\end{eqnarray}
The orders of the remaining variables can be found as indicated in Prop.\ref{prop1}:
\begin{eqnarray}
a_{15} &=& a_{12} + \gamma_{26}  - \gamma_{27}, \\
a_{16} &=& a_{13} + \gamma_{28}  - \gamma_{29}, \\
a_{11} &=& a_{12} + a_{13} + \gamma_{22} + \gamma_{35} - \gamma_{30}.
\end{eqnarray}
The polyhedron of tropical solutions defined by Eq.\eqref{pol} can be partitioned into
minimal branches (also polyhedra). This can be done by checking which term is
dominant in the ordinary differential equations for the variables $x_{12}$, $x_{13}$
and $x_{14}$ (see Eqs.\eqref{system}).
The result is that there are at most four minimal branches defined
by one of the conditions
\begin{eqnarray}
\{ a_{12} + \gamma_{26} < a_{12} + a_{13} + \gamma_{22} + \gamma_{35} \} \cap
\{ a_{13} + \gamma_{28} < a_{12} + a_{13} + \gamma_{22} + \gamma_{35} \}  \notag \\
\{ a_{12} + \gamma_{26} < a_{12} + a_{13} + \gamma_{22} + \gamma_{35} \} \cap
\{ a_{13} + \gamma_{28} > a_{12} + a_{13} + \gamma_{22} + \gamma_{35} \}  \notag \\
\{ a_{12} + \gamma_{26} > a_{12} + a_{13} + \gamma_{22} + \gamma_{35} \} \cap
\{ a_{13} + \gamma_{28} < a_{12} + a_{13} + \gamma_{22} + \gamma_{35} \} \notag \\
\{ a_{12} + \gamma_{26} > a_{12} + a_{13} + \gamma_{22} + \gamma_{35} \}
 \cap  \{ a_{13} + \gamma_{28} > a_{12} + a_{13} + \gamma_{22} + \gamma_{35} \} \label{eq:branches}
 \end{eqnarray}

\begin{figure}
\includegraphics[width=1.0\textwidth]{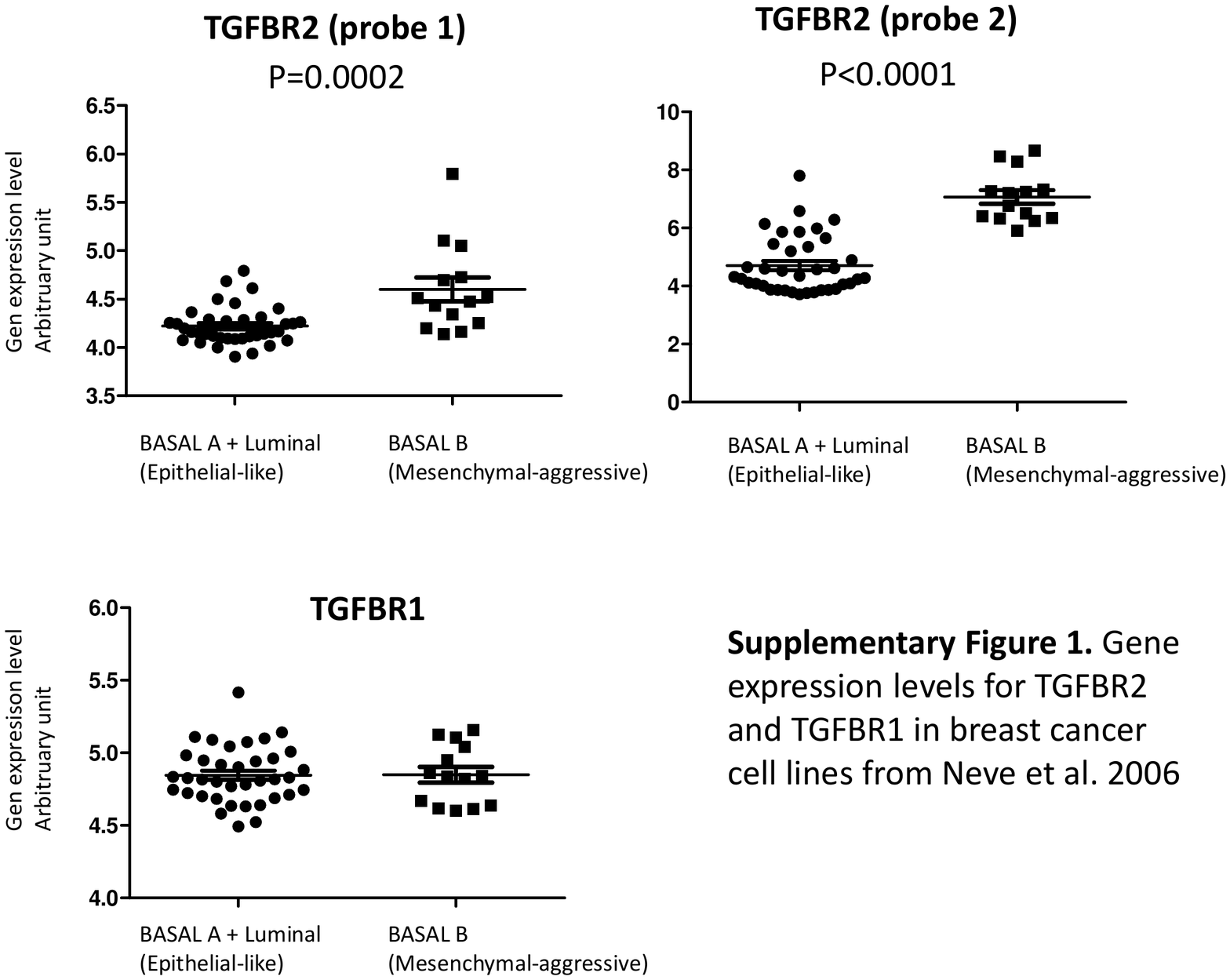}
\end{figure}

\end{document}